\documentclass[aps,pre,reprint,superscriptaddress,nofootinbib]{revtex4-1}
\usepackage{graphicx}
\usepackage{amsmath}
\usepackage{amsfonts}
\usepackage{bm}
\usepackage{hyperref}
\usepackage{xcolor}

\hypersetup{colorlinks=true,citecolor=black,linkbordercolor=red,linkcolor=blue}

\usepackage{url,lineno,microtype,color}

\usepackage[]{algorithm2e}

\newcommand{\DAB}[1]{#1}
\newcommand{\DMS}[1]{#1}
\newcommand{\Q}{\mathbf{Q}}
\newcommand{\tr}{\textrm{tr}}
\def\PackageName{\textit{openQmin}}

\begin{document}

\title{Fast, scalable, and interactive software for Landau-de Gennes numerical modeling of nematic topological defects} 

\author{Daniel M. Sussman}
\email{daniel.m.sussman@emory.edu}
\affiliation{Syracuse University, Department of Physics, Syracuse, NY, USA}
\affiliation{Emory University, Department of Physics, Atlanta, GA, USA}
\author{Daniel A. Beller}
\email{dbeller@ucmerced.edu}
\affiliation{University of California, Merced,  Department of Physics, Merced, CA, USA}

\date{\today}

\begin{abstract}
Numerical modeling of nematic liquid crystals using the tensorial Landau-de Gennes (LdG) theory provides detailed insights into the structure and energetics of the enormous variety of possible topological defect configurations that may arise when the liquid crystal is in contact with colloidal inclusions or structured boundaries. However, these methods can be computationally expensive, making it challenging to predict (meta)stable configurations involving several colloidal particles, and they are often restricted to system sizes well below the experimental scale. Here we present an open-source software package that exploits the embarrassingly parallel structure of the lattice discretization of the LdG approach. Our implementation, combining CUDA/C++ and OpenMPI, allows users to accelerate simulations using both CPU and GPU resources in either single- or multiple-core configurations. We make use of an efficient minimization algorithm, the Fast Inertial Relaxation Engine (FIRE) method, that is well-suited to large-scale parallelization, requiring little additional memory or computational cost while offering performance competitive with other commonly used methods. In multi-core operation we are able to scale simulations up to supra-micron length scales of experimental relevance, and in single-core operation the simulation package includes a user-friendly GUI environment for rapid prototyping of interfacial features and the multifarious defect states they can promote. To demonstrate this software package, we examine in detail the competition between curvilinear disclinations and point-like hedgehog defects as size scale, material properties, and geometric features are varied. \DAB{We also study the effects of an interface patterned with an array of topological point-defects.}
\end{abstract}

\maketitle

\section{Introduction}

Nematic liquid crystals' combination of fluidity and orientational order   both underlies nematics' widespread technological applications and endows them with topological defects, localized breakdowns in the orientational order stabilized by the medium's broken symmetries. The topological defects of nematics have been integral to the study of liquid crystals since the field's infancy \cite{Friedel1922}. 

Besides their role as tabletop physical realizations of profound topological ideas, nematic topological defects \DAB{-- including disclination lines, point-like hedgehogs, and surface-bound boojums --} are of great interest for their importance in nematic colloidal suspensions  \cite{poulin1997novel}.
 These composite materials, formed by suspensions of colloidal particles or nanoparticles in nematics, promise new routes to directed self-assembly and self-organization. Nanoparticles in nematics are pushed by elastic forces to assemble in pre-existing defect lines, meaning that sculpted disclinations provide a path to controlled nanoparticle assembly.  Applications include  plasmonic properties for metamaterials  \citep{dickson2008electronically,liu2010self}, molecular self-assembly  \cite{wang2016topological}, and quantum-dot assembly in microshells \cite{rodarte2013quantum,rodarte2015quantum}.  Even greater complexity arises in the cases of colloidal particles in the size range of tens of nanometers to several microns, which often have companion topological defects and which interact through forces mediated by nematic elasticity. Self-assembled structures of colloidal particles with companion defects include bound pairs, chains \citep{poulin1997novel,Musevic2006} and triclinic 3D crystals \citep{mundoor2016triclinic};  with the aid of laser tweezers, other configurations such as  3D crystals with tetragonal symmetry \cite{nych2013assembly} and sophisticated disclination knots   \citep{Ravnik2007,Tkalec:2011lj,tasinkevych2014splitting,machon2014knotted} can be stabilized. Tailored self-assembled colloidal structures hold promise as optical metamaterials for photonics applications, such as photonic bandgap crystals and microlasers \citep{Ravnik2011,lavrentovich2011liquid,humar20103d,muvsevivc2019interactions}. 
 
\DAB{Nematic defect configurations can also be controlled by nontrivial boundary surfaces \cite{serra2016curvature}. Substrate patterning strategies include  topographic variations such as ``lock-and-key'' docking sites for colloidal particles \cite{hung2007nanoparticles,eskandari2014particlesoftmatter,luo2016experimental} and   chemical patterning where the boundary condition shifts abruptly \cite{guillamat2014electric,kos2015relevance}. 
Complex director fields, including disclinations, can be prescribed on a substrate by photoalignment \cite{peng2015liquid} or by scribing with an atomic force microscope \cite{murray2014creating}. Confinement in geometries such as capillaries \cite{williams1973screw}, droplets \cite{volovik1983topological}, shells \cite{nelson2002toward}, and thin films \cite{Lavrentovich1990} produces a wealth of point- and line-defect behaviors stabilized by topology and energetics.}

The rapidly expanding variety of experimentally created \DAB{nematic defect} configurations has benefited greatly from the understanding provided by robust modeling approaches.  One set of approaches is based on the Frank-Oseen elastic free energy, which penalizes deformations of the nematic director $\hat n(\mathbf{x})$, and which in its simplest form reads
\begin{equation}\label{eq: FrankOseen}
\mathcal F_{\mathrm{FO}}^{(1)} = \frac{K}{2} \int dV \sum_{i,j=1}^3 (\partial_i n_j)^2. 
\end{equation}
The superscript $(1)$ refers to the approximation of a single elastic constant $K$ in this expression. However, the $\hat n = - \hat n $ symmetry of nematics presents challenges for this model in the presence of disclinations with half-integer winding number, especially if their locations are not known beforehand.  

This difficulty is resolved by the Landau-de Gennes (LdG) model, the theoretical approach which is the focus of this work and which we review in Sec.~\ref{sec: LdGreview}.  The LdG framework takes as its order parameter the  second-rank traceless nematic order tensor $Q_{ij}(\mathbf{x})$, and is well-suited to modeling arbitrary disclination configurations, as well as biaxial nematics and the blue phases \cite{Ravnik2011,wright1989crystalline}.  While little is known analytically about free energy minimizers in LdG theory in \DAB{any but the simplest geometries} \cite{alama2016minimizers,alama2016analytical}, numerical minimization of the LdG free energy has been fruitfully applied over a wide range of systems \citep{nych2013assembly, Ravnik2007,Tkalec:2011lj,Ravnik2011,Kralj1992,tkalec2008interactions,emervsivc2019sculpting, luo2018tunable, tasinkevych2014dispersions, hashemi2015equilibrium, hung2009faceted, HungPRE2009, hung2006anisotropic,  beller2015shape, araki2006colloidal, Skarabot2008, eskandari2014particlesoftmatter, vskarabot2008hierarchical,  ravnik2009landau,Mori1999}. Additionally, flow dynamics of nematics, including active nematic systems, can be modeled by supplementing the LdG free energy with hydrodynamical equations \DAB{as formulated by Beris and Edwards \cite{beris1994thermodynamics} or by Qian and Sheng \cite{qian1998generalized} and solved by methods such as lattice Boltzmann \cite{denniston2001lattice,spencer2006lattice,2019arXiv190601129C,marenduzzo2007steady,cates2009lattice},  multiparticle collision dynamics and related off-lattice methods \cite{lee2015stochastic,shendruk2015multi,mandal2019multiparticle}, or finite difference and finite element approaches \cite{james2008computer,giomi2011excitable}. Some methods incorporate a fast relaxation of the momentum compared to the director,  to account for the separation in time scales for these relaxations in typical molecular liquid crystals \cite{svenvsek2002hydrodynamics,spencer2006lattice}.}

The broad usefulness of the LdG theory goes hand in hand with a significant limitation of scale: Resolving defects at \textit{a priori} unknown locations requires the simulation lattice spacing to be comparable to or smaller than the size of the defect core, the region in which nematic order breaks down, which in thermotropic nematics is typically a few nanometers. This is often thousands of times smaller than the individual micron-scale colloidal particles of interest. Therefore, a faithful rescaling of the experimental system in numerics would require of order at least $10^9$ lattice sites even for configurations involving only a small number of such colloids.

Accessing such experimentally relevant lattice sizes presents computational challenges not often seen in the simulations of glassy and polymeric soft matter systems. The demands on system memory quickly become prohibitive: simply maintaining the five independent degrees of freedom at each lattice site and storing the necessary change in those variables from one minimization step to the next at $10^9$ lattice sites requires 80 GB -- more than on most current commodity desktops and larger than the memory capacity of any CUDA-capable GPU\footnote{As of July, 2019}. Additionally, there is a large  direct computational cost of even simple manipulations acting on so many degrees of freedom; this contributes to the significant wall-time required for most numerical energy minimizations and presents challenges for efficient exploration of parameter spaces and colloidal particle positions.

Consequently, LdG numerical modeling is typically applied to systems significantly scaled down, with respect to a fixed defect core size, as compared with the experiments that they are intended to model.   While important qualitative insights about defects and director fields can often be obtained by scaling down the experimental dimensions, the change in size ratios makes quantitative prediction challenging. There can also be major qualitative differences. The most well-known of these is the form of the companion defect to a  particle with homeotropic (normal) anchoring: Micron-scale particles typically have hyperbolic hedgehog companions (in the absence of confinement or external fields) \citep{poulin1997novel}, whereas particles in the few hundred nm or smaller size range have disclination loops in the ``Saturn ring'' configuration \citep{terentjev1995disclination,Stark2001}. This constitutes a major challenge in modeling systems with multiple colloid-hedgehog pairs. Experimental work on high-aspect ratio colloidal particle shapes observes both hedgehogs and disclination loops, but numerical modeling has been limited to the line defect case \cite{tkalec2008interactions,gharbi2013microbulletPublished,luo2019deck,luo2019deck,hung2006anisotropic,hung2009faceted,HungPRE2009,tasinkevych2014dispersions,beller2015shape,hashemi2015equilibrium}.   Adaptive mesh refinement  in finite-element simulations can help to avoid computational and memory expense in regions not near defects \citep{bajc2016mesh} but typically does not remove the need to scale down.

In this work we present an open-source finite difference-based implementation of LdG free energy minimization with nontrivial boundary conditions, ``\PackageName'' \citep{landauDeGUI}, using a combination of approaches designed to address the challenges described above.  It is written for heterogeneous CPU and GPU operation  to target two complementary research goals. First, it offers a user-friendly GUI environment for rapid prototyping of topological defect configurations as a function of liquid crystal parameters, boundary geometry, and the presence of colloidal inclusions. Simultaneously, it targets large-scale systems using OpenMPI \citep{gabriel2004open} to support parallelization across both CPU and GPU resources to scale up to the supra-micron length scales of experimental relevance. We employ efficient minimization algorithms, such as the Fast Inertial Relaxation Engine (FIRE) method, to maintain reasonable convergence times even for large-scale parallelized calculations. 

The remainder of the paper is structured as follows. We begin with a review of the LdG theory in Section \ref{sec: LdGreview}. Section \ref{sec: NumericalApproach} lays out our numerical approach, first discretizing the LdG theory for a finite-difference method, and then outlining our use of minimization algorithms and OpenMPI parallelization. \DMS{In Sec. \ref{sec:SampleStudy} we present two sample studies demonstrating the effectiveness of this approach. We first perform a classic analysis of the companion defects to homeotropic spherical particles at varying system sizes, and then examine the effects of a boundary patterned with surface disclinations in a supra-micron-scale system.} Section \ref{sec: GUI} briefly describes the GUI version of \PackageName\ with an example of the rapid prototyping workflow it enables. Finally, in Section \ref{sec: Discussion}, we discuss both the range of use we foresee for \PackageName\  and some future directions for additional physics that could be studied in this framework.

\section{Landau de-Gennes theory for nematic liquid crystals }\label{sec: LdGreview}
Here we give a brief overview of those aspects of the LdG theory used in our numerical approach. The theory is of course well-established \cite{deGennes_book} and its use in a finite difference numerical free energy minimization scheme is described in several sources; the reader is directed to  Ref.~\citep{ravnik2009landau} for a thorough explanation.

Uniaxial nematic liquid crystals are characterized by orientational ordering of nematogens (molecules or suspended anisotropic particles) along a director, $\hat{n}$, which is a unit vector with the identification $\hat{n}=-\hat{n}$.To respect that symmetry consistently, which is important at disclinations of half-integer winding number, we take as order parameter not a director but a second-rank tensor. 
This is the traceless, symmetric tensor field  $\Q(\mathbf{x})$, whose lattice discretization is the fundamental object of the LdG modeling approach.  $\Q$  is related to $\hat n $ by \citep{mottram2014introduction}
\begin{eqnarray}\label{eq:Qtensor}
Q_{\alpha\beta} &=& \frac{3}{2}S\left(n_\alpha n_\beta-\frac{1}{3}\delta_{\alpha\beta}\right) +\frac{1}{2} S_B(m_\alpha m_\beta - l_\alpha l_\beta).
\end{eqnarray}
Here, $S$ is the degree of uniaxial nematic order, and $S_B$ is the degree of biaxial order distinguishing a preferred direction $\hat m\equiv - \hat m $, perpendicular to $\hat n$, from $\hat l \equiv \hat n \times \hat m$. The nematic director can be recovered as the eigenvector corresponding to the largest eigenvalue of $\Q$, which equals $S$. Most nematics are unaxial, so the equality $S_B=0$ is true in the absence of distortions and represents a good approximation away from defects. In this uniaxial limit, Eq. \ref{eq:Qtensor} reduces to 
\begin{equation}
Q_{\alpha\beta}  = \frac{3}{2}S\left(n_\alpha n_\beta-\frac{1}{3}\delta_{\alpha\beta}\right). 
\end{equation}

\subsection{Phenomenological free energy density}
The LdG theory constructs a phenomenological  free energy $\mathcal F$ as a functional of $\Q(\mathbf{x})$. We can write this functional schematically as \citep{ravnik2009landau,mottram2014introduction}:
\begin{align}\label{eq:Fldg}
\mathcal{F}[\Q] &=& \int_V \left( f_{\mathrm{bulk}}+f_{\mathrm{distortion}} +f_{\mathrm{external}} \right) dv \\
&&+  \sum_\alpha \int_{S_\alpha }\left( f_{\mathrm{boundary}}^\alpha \right) ds\nonumber
\end{align}
The first integral, over the volume of the nematic, has three  free energy density terms incorporating respectively the energetic costs arising from  deviations away from the thermodynamically preferred degree of nematic order $S=S_0$, from elastic distortions, and from  external fields. The second integral, summing over all boundary surfaces $S_\alpha$ in contact with the nematic, incorporates the anchoring energy associated with each interface, including the surfaces of colloidal particles; its form may be different for different surfaces. We address each component in turn:

\subsubsection{Bulk free energy}
The first free energy density term in Eq.~\ref{eq:Fldg} gives a Landau free energy for the isotropic-nematic phase transition, written in terms of rotational invariants of $\Q$ in a Taylor expansion about the isotropic, $\Q=0$ state \citep{Schophol1987}:
\begin{equation}\label{eq:fphase}
f_{\mathrm{bulk}} = \frac{A}{2} \tr\left(\Q^2\right)+\frac{B}{3} \tr\left(\Q^3\right)+\frac{C}{4} \left(\tr\left(\Q^2\right)\right)^2.
\end{equation}
The parameter $A\propto (T-T_{NI}^*)$, where $T_{NI}^*$ is the temperature at which the isotropic phase is destabilized. In the uniaxial limit $f_{\mathrm{bulk}}$ becomes a polynomial in the degree of order,
\begin{equation} \label{eq:fphaseS}
f_{\mathrm{bulk}} = \frac{3}{4} A S^2  + \frac{1}{4} B S^3 + \frac{9}{16} C S^4,
\end{equation}
which is minimized either  by $S=0$ or by 
\begin{equation} \label{eq:S0def}
S=S_0\equiv \frac{-B+\sqrt{B^2-24AC}}{6C}.
\end{equation}
In the nematic phase, the absolute value of 
\begin{equation}\label{eq:f0}
f_0 \equiv f_{\mathrm{bulk}}(S=S_0)
\end{equation}
provides a free energy penalty per unit volume to the melted cores of defects, where $S\rightarrow 0$.

\subsubsection{Distortion free energy}
The distortion free energy density models the elasticity of the nematic phase, and represents the LdG counterpart to the Frank-Oseen free energy density. The latter, in full generality up to second derivatives of $\hat n$, is 
\begin{eqnarray} 
f_{\mathrm{FO}} &=& \frac{1}{2} \bigl\{K_1(\nabla  \cdot \hat n)^2 + K_2 (\hat n \cdot (\nabla \times \hat n) +  q_0)^2 + K_3  |(\hat n \cdot \nabla )\hat n |^2   \nonumber   \\
& & \quad  + K_{24} \nabla \cdot [ (\hat n \cdot \nabla) \hat n  - \hat n (\nabla \cdot \hat n ) ] \bigr\}.  \label{eq: FOfull}
\end{eqnarray}
The parameters in this expression are the splay ($K_1$), twist ($K_2$), bend ($K_3$), and saddle-splay ($K_{24}$) elastic constants, and the spontaneous chiral wavenumber $q_0$ which is nonzero in the cholesteric and blue phases. \DAB{Other common  conventions for the saddle-splay energy density replace $K_{24}$ in Eq.~\ref{eq: FOfull} by either $2K_{24}$ or $2(K_2+K_{24})$.} Eq.~\ref{eq: FOfull} reduces to Eq.~\ref{eq: FrankOseen} under the ``one-constant approximation'' $K_1 = K_2 = K_3 = K_{24} \equiv K $ and $q_0=0$.  \DAB{The one-constant approximation is a reasonable simplification for  many molecular liquid crystals, where $K_1$, $K_2$, and $K_3$ typically differ by less than a factor of 5  \cite{deJeu1976Determination,Madhusudana1982Elasticity}.}

The most general form of $f_{\mathrm{distortion}}$ that we employ, following Refs.~\cite{poniewierski1985free,Mori1999, mottram2014introduction},  includes all gradient terms of quadratic order in $\Q$ allowed by symmetry, plus one term at cubic order: 
\begin{eqnarray} 
f_{\mathrm{distortion}} &=& \frac{L_1}{2}\frac{\partial Q_{ij}}{\partial x_k}\frac{\partial Q_{ij}}{\partial x_k}   + \frac{L_2}{2}\frac{\partial Q_{ij}}{\partial x_j}\frac{\partial Q_{ik}}{\partial x_k}  \nonumber \\
&\quad\quad+& \frac{L_3}{2}\frac{\partial Q_{ik}}{\partial x_j}\frac{\partial Q_{ij}}{\partial x_k}+\frac{L_4}{2}\epsilon_{lik}Q_{lj}\frac{\partial Q_{ij}}{\partial x_k} \nonumber \\
&\quad\quad+& \frac{L_6}{2} Q_{lk} \frac{\partial Q_{ij}}{\partial x_l}\frac{\partial Q_{ij}}{\partial x_k}, \label{eq:fdistortionFull}
\end{eqnarray}
where Einstein summation over repeated indices is implied, and $\epsilon$ is the Levi-Civita tensor. Equation~\ref{eq:fdistortionFull}  corresponds in the uniaxial limit to Eq.~\ref{eq: FOfull} with the identifications \cite{Mori1999}
\begin{eqnarray*}
L_1 &=& \frac{2}{27 S^2} \left(K_3-K_1+3 K_2 \right), \\
L_2 &=& \frac{4}{9S^2} \left(K_1 - K_{24} \right), \\
L_3 &=& \frac{4}{9S^2} \left( K_{24} - K_2 \right), \\
L_4 &=& - \frac{8}{9 S^2} q_0 K_2, \\
L_6 &=& \frac{4}{27S^3} (K_3 - K_1 ). 
\end{eqnarray*} 

The one-constant approximation in the absence of spontaneous chiral ordering sets $L_2=L_3=L_4=L_6=0$, leaving the much simpler and more computationally efficient form 
\begin{equation}\label{eq:fd1}
f_{\mathrm{distortion}}^{(1)} = \frac{L_1}{2} \frac{\partial Q_{ij}}{\partial x_k}\frac{\partial Q_{ij}}{\partial x_k},
\end{equation}
which corresponds in the uniaxial limit to Eq.~\ref{eq: FrankOseen} with $L_1 = 2/(9S^2) K$. 

Taking this simpler form of the distortion energy density, we estimate the defect core size by considering a distorted uniaxial nematic configuration at $S=S_0$ with $\hat n$ varying with typical gradient $1/\ell$. Roughly speaking, the energy well depth $f_0$ (Eq.~\ref{eq:f0}) gives the threshold value for $f_{\mathrm{distortion}}$ at which distortions become so energetically costly that a local melting of nematic order occurs instead. This length  $\ell = \xi_N$, the nematic correlation length (or coherence length), sets the size of the defect core: 
\begin{equation}\label{eq:xiN}
\xi_N \sim \sqrt{L_1/|f_0|}. 
\end{equation}

\subsubsection{External fields free energy}

The response of the nematic to an external magnetic field $\mathbf{H}$ or an external electric field $\mathbf{E}$ is modeled by the free energy density term
\begin{equation}
f_{\mathrm{external}} = - \tfrac{1}{3} \mu_0 H_i \Delta \chi Q_{ij} H_j  - \tfrac{1}{3}  \varepsilon_0 E_i \Delta \varepsilon Q_{ij} E_j 
\end{equation}
where $\Delta \chi$ and $\Delta \varepsilon$ are the anisotropic parts (difference in principal values corresponding to $\hat n$ and its perpendicular directions) of the magnetic susceptibility tensor and dielectric tensor, respectively \cite{Kralj1992}, and $\mu_0$ and $\varepsilon_0$ are respectively the magnetic permeability and electric permittivity of free space. (We omit here the terms for the isotropic parts of these tensors, as they do not couple to $\Q$.) In the uniaxial limit, the right-hand side becomes 
\DAB{$-\frac{1}{2} S \mu_0  \Delta \chi  (\mathbf{H}\cdot \hat n )^2  - \frac{1}{2}  S\varepsilon_0 \Delta \varepsilon (\mathbf{E}\cdot \hat n )^2$}
 (again dropping isotropic terms with no coupling to $\hat n $). Positive $\Delta \chi$ or $\Delta \varepsilon$ will favor alignment of $\hat n $ with $\mathbf{H}$ or $\mathbf{E}$. 

\subsubsection{Boundary free energy}

Boundary surfaces, including the surfaces of embedded colloidal particles, generally impose an anchoring surface energy density $f_{\mathrm{boundary}}$ representing the surface tension's dependence on the director at the surface. In terms of the director, a common modeling choice for  the anchoring energy is the Rapini-Papoular form $-\tfrac{1}{2} W^\alpha_{\mathrm{RP}} (\hat \nu^\alpha \cdot \mathbf{n})^2$ where $\hat \nu^\alpha$ is the surface normal vector  and $|W^\alpha|$ is the anchoring strength of surface $\alpha$ \cite{rapini1969distorsion}. Homeotropic (normal) anchoring follows from $W_{\mathrm{RP}}>0$, whereas $W_{\mathrm{RP}}<0$ creates degenerate planar anchoring, which equally favors every direction perpendicular to $\hat \nu^\alpha$. The same anchoring functional can favor a different anchoring direction, for example an in-plane direction in the case of oriented planar anchoring, using $W_{\mathrm{RP}}>0$ with the replacement of $\hat \nu^\alpha$ by the favored direction. 

In LdG theory, for homeotropic or other oriented anchoring, the Rapini-Papoular form is generalized as the Nobili-Durand surface anchoring form \cite{nobili1992disorientation},
\begin{equation}
f_{\mathrm{boundary}}^\alpha  = W_{\mathrm{ND}}^\alpha \tr\left( (\Q-\Q^0)^2\right) = W_{\mathrm{ND}}^\alpha (Q_{i j} - Q^\alpha_{ i j})(Q_{i j} - Q^\alpha_{ i j}),
\end{equation}
where $W_{\mathrm{ND}}^\alpha>0$ is the anchoring strength of surface $\alpha$ and  the surface-preferred $\Q$-tensor, $\Q^\alpha$, is usually taken to be  $Q^\alpha_{i j} = \frac{3}{2} S_0 (n_i^\alpha n_j^\alpha - \tfrac{1}{3} \delta_{ij})$, with $\hat n ^\alpha = \hat \nu^\alpha$ or some other surface-preferred director. 

For degenerate planar anchoring, the Nobili-Durand form is not suitable, and we use instead the following free energy due to Fournier and Galatola \cite{Fournier2005}:
\begin{equation}\label{eq:planarDegenerate}
f_{\mathrm{boundary}}^\alpha  = W_{\mathrm{FG}}^\alpha (\tilde{Q}_{ij}-\tilde{Q}^\bot_{ij})(\tilde{Q}_{ij}-\tilde{Q}^\bot_{ij}) ,\end{equation}
where $\tilde{Q}_{ij} = Q_{ij}+S_0\delta_{ij}/2$ and $\tilde{\Q}^\bot$ is the projection onto $\hat{\nu}^\alpha $ via $\tilde{Q}^\bot_{ij} = P_{ik} -\tilde{Q}_{k\ell} P_{\ell j}$ for $P_{ij} = \delta_{ij} -  \nu_i^\alpha \nu_j^\alpha$. \DAB{A}ssuming $\Q$ is uniaxial with $S=S_0$, the Rapini-Papoular anchoring is recovered with $W_{\mathrm{RP}}^\alpha =9  S_0^2 W_{\mathrm{ND,FG}}^\alpha$. 

\section{Numerical approach} \label{sec: NumericalApproach}
\subsection{Overview}
The primary contribution of this work is the presentation of an open-source numerical implementation that exploits the embarrassingly parallel structure of the lattice discretization of the above phenomenological theory.  Our implementation, combining CUDA/C++ \citep{nvidia2011nvidia} and OpenMPI \citep{gabriel2004open}, was written with extreme flexibility in mind to allow users to accelerate simulations large and small using combinations of available CPU and GPU resources in either single- or multiple-core configurations.

The foundation of the software package, ``dDimensionalSimulation,'' is a set of generic classes meant to execute simulations of $N$ interacting units, each consisting of $d$ scalar degrees of freedom, using data structures appropriate for efficient execution on either CPU or GPU resources. These generic classes serve as the template for \emph{models} which instantiate the $dN$ total degrees of freedom, \emph{forces} which compute interactions between degrees of freedom, \emph{updaters} which can change the degrees of freedom (e.g., by implementing equations of motion), and \emph{simulations} which tie objects of these various types together. The present work focuses on implementing the details of these classes to carry out lattice-based LdG modeling to find energy-minimized configurations of equilibrium nematics in the presence of various boundary conditions. The general structure we have employed was chosen to allow flexibility in future development, for example to derive new model classes which include not only the $\Q$-tensor but also density and velocity degrees of freedom, as would be appropriate for modeling active nematic systems \citep{2019arXiv190601129C,marenduzzo2007steady,cates2009lattice,giomi2011excitable}. 

\begin{figure}[h!]
\begin{center}
\includegraphics[width=15cm]{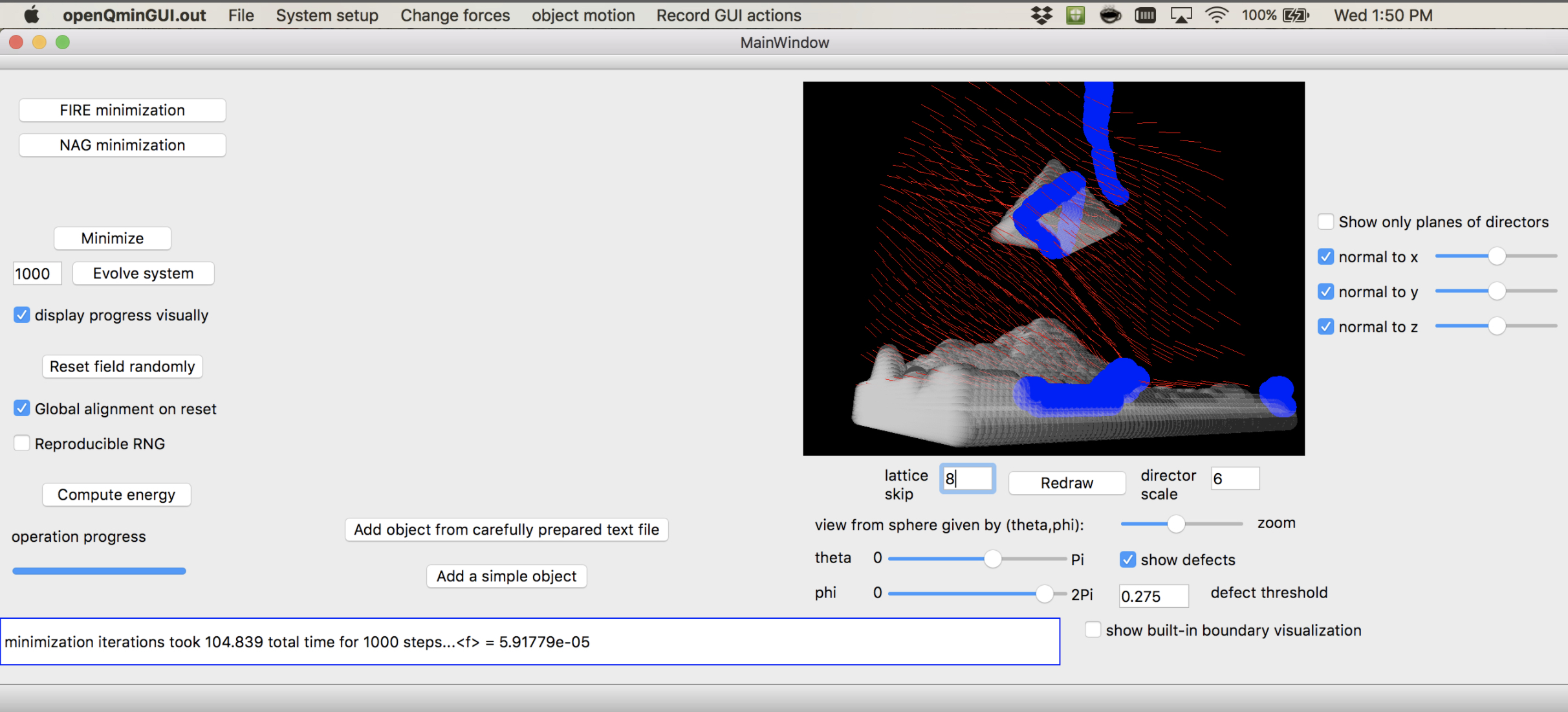}
\end{center}
\caption{Snapshot of the graphical user interface provided by \PackageName, here shown simulating the defect structure near a pyramidal colloid above a topographically nontrivial boundary, all with oriented anchoring along user-specified directions to approximately model homeotropic surfaces.}\label{fig:gui}
\end{figure}

In addition to writing efficient code to carry out the required lattice-based minimizations of the $\Q$-tensor field in a domain, we also advocate the use of the graphical user interface (GUI) we developed to rapidly prototype and explore the effects of particular boundaries, colloidal inclusions, and external fields that may be of experimental interest. The GUI allows a wide variety of user operations -- adding boundary objects at any stage of the simulation, starting and stopping minimization, adding or removing external fields at will --  all while visualizing the resulting defect structure and recording configurational details. A snapshot of the GUI is shown in Fig. \ref{fig:gui}, and more details of the available features are given in Sec. \ref{sec: GUI}. We envision that this capability will allow for rapid prototyping of experimental geometries in the search for particular controllable defect states; running on a single GPU allows real-time visualization of lattices in the low-millions of total sites. We have also included several example files that use the code in a non-GUI mode; these can then use OpenMPI to parallelize across either CPU or GPU resources to scale up to lattices that represent supra-micron-scale liquid crystal systems. 

\onecolumngrid
 
\twocolumngrid

\subsection{Lattice discretization and energy minimization}

The finite difference lattice calculations employed in this work use a regular cubic lattice discretization of space, with a $\Q$-tensor defined at each site $\vec{x}=\{x,y,z\}$. The lattice $\Delta x$ spacing can be related to physical quantities through a natural nondimensionalization of the free energy density, $\tilde f\equiv f/|A|$, which  implies a nondimensionalization of  the elastic constants $\tilde L_i\equiv L_i/(|A| \Delta x^2)$. In the one-constant approximation, we thus have $\Delta x^2= L_1/(\tilde L_1 |A|)$.
In this work we set $\tilde L_1 = 2.32$. To model 5CB, following Ref.~\cite{ravnik2009landau} we take $A=-0.172\times10^6\,\mathrm{J}/\mathrm{m}^3$, $B=-2.12\times10^6\,\mathrm{J}/\mathrm{m}^3$, $C=1.73\times10^6\,\mathrm{J}/\mathrm{m}^3$, and $K=L_1 \cdot 9 S_0^2/2=1\times10^{-11}\,\mathrm{N}$ where $S_0\approx 0.53$. These give a lattice spacing of $\Delta x \approx 4.5\,\mathrm{nm}$, which is at the few-nm scale of the defect core in 5CB. Note that the nondimensionalization of all constants by an energy scale $|A|$ and a length $\Delta x$ is implicitly made for all values in \PackageName, including in the GUI.

The symmetry and tracelessness of $\Q$ leaves five independent degrees of freedom, which we take to be $\vec {q} \equiv \{Q_{xx}, Q_{xy}, Q_{xz}, Q_{yy}, Q_{yz}\}$ at each of the $N$ lattice sites in the simulation domain. We write the local free energy density $f(\vec{x})$ in terms of these five independent variables, so that the symmetry and tracelessness of  $\Q$ are automatic (rather than being maintained by projection operations following update steps \citep{ravnik2009landau}). We also label each site with an integer ``type,'' indicating whether it is a bulk site, a boundary site, or a site inside an object (for instance, the interior of a colloidal inclusion, or part of a bounding surface), depending on the geometry of the problem. Only bulk and boundary sites are ``simulated sites'', meaning $\Q$ is defined there.

We discretize the total free energy, $\mathcal{F} = \sum_{i=1}^N f(\vec{x}_i)$, using a finite-difference approach over the $5N$ independent variables. For the distortion terms we allow the user to select either the more general expression, Eq. \ref{eq:fdistortionFull}, or the more computationally efficient expression of Eq. \ref{eq:fd1}. For the terms in $f_{\mathrm{distortion}}$ which contain spatial first derivatives of $\Q$, we consider first-order forward and backward finite difference approximations,
\begin{equation}
\left(\frac{\partial Q_{ij} }{\partial x_k}\right)(\vec{x}) \approx  \begin{cases}  Q_{ij} (\vec{x} + \hat x_k ) - Q_{ij} (\vec {x}) & \text{(forward})  \\ Q_{ij}(\vec{x}) - Q_{ij} (\vec{x} - \hat x_k) & \text{(backward)\DAB{.}} \end{cases} \label{eq: onesidedderivatives}
\end{equation}
Here $\hat x_k$ is the unit vector in the $x_k$ direction, and $\vec{x}$ is the site where the calculation is taking place. \DAB{The choice of a regular cubic lattice makes these derivative approximations straightforward to calculate.} The forward and backward finite differences are each compatible with the simulation domain only if ($\vec{x} + \hat x_k$), $(\vec{x} - \hat x_k)$, respectively, are simulated (bulk or boundary) sites. We then take, as the discretized expression of $f$, Eq. \ref{eq:Fldg} averaged over all forward and backward finite difference expressions for each of $k=1,2,3$ that are allowed by the geometry of the simulation domain. A bulk site, therefore, has a local free energy averaged over $2^3$ such combinations, while a boundary site has fewer. We use these averages over different expressions for the finite differences, rather than using a single centered finite difference formula ($\partial Q_{ij} /\partial x)(\vec{x}) \approx \tfrac{1}{2} [Q_{ij}(\vec{x}+\hat x_k) - Q_{ij}(\vec{x} - \hat x_k)]$, because \DAB{using the latter form in Eq.~\ref{eq:fd1} produces no terms coupling $Q_{ij}(\vec{x})$ to its nearest neighbors, of the form $Q_{ij}(\vec{x}) Q_{ij}(\vec{x} \pm \hat x_k)$. This use of the centered first derivative expression would therefore create an artificial (and undesirable) lattice doubling effect in our approach, with even sites and odd sites evolving independently. For curved boundaries such as on spherical colloidal particles, well-known inaccuracies are introduced in the finite difference calculations by the discretization of boundaries as sites in the cubic lattice \cite{noye1990accurate}. Specifically, errors of order $O(\Delta x)$ in $Q_{ij}(\vec{x})$ are introduced, leading to truncation error of $O(1)$ (which do not diverge as the lattice spacing is refined) in the first derivative approximations of Eq.~\ref{eq: onesidedderivatives}. }

Finally, we minimize $\mathcal{F}$ as a cost function over the $5N$ independent variables $q_i(\vec{x}_j)$, $i=1,\dots,5$, $j=1,\dots,N$. The gradient of $\mathcal{F}$ in this $5N$-dimensional space is calculated by explicitly differentiating the expression for $\mathcal{F}$ with respect to each $q_i(\vec{x}_j)$ degree of freedom. This explicit differentiation of a cost function is an alternative to the approach of analytically deriving local forces (molecular fields) from the Euler-Lagrange equations, projecting to recover symmetry and tracelessness, and then discretizing those expressions. \DAB{While the Euler-Lagrange equations have separate forms for the bulk and the boundaries, in the approach used here forces are derived from the cost function in formally the same way for bulk and boundary sites.}

\DMS{We emphasize that by discretizing space, we can directly map the problem of solving the LdG partial differential equations to finding the minima of a complex energy landscape (where the $\Q$-tensors on each lattice site are the degrees of freedom). For instance, many PDE solvers implement steepest descent relaxation, which can be directly interpreted as overdamped molecular dynamics at zero temperature. This allows us to turn to the wealth of} existing algorithmic approaches from the field of nonlinear optimization, including minimization techniques such as quasi-Newton methods, (conjugate) gradient descent, and momentum-based techniques such as Nesterov's accelerated gradient \citep{nocedal2006numerical}. Since our aim is to be able to scale up to large systems, we ignore minimizers which require second-order derivatives of the cost function, and we find that even limited-memory quasi-Newton methods such as L-BFGS impose too-strong a memory requirement for \DMS{many of} our purposes. Additionally, while conjugate gradient is appealing in having only marginal extra memory requirements and being much faster than simple gradient descent, it involves frequent line searches that require expensive repeated evaluations of the free energy density and imposes additional parallelization costs.

Thus, although we have implemented many of the above-named minimizers in \PackageName, we 
focus our attention on the use of the Fast Inertial Relaxation Engine (FIRE) method of energy minimization \citep{bitzek2006structural}. FIRE falls into the class of ``gradient plus momentum''-style minimization algorithms, and it additionally rescales the ``velocity'' \DMS{(fictitious additional variables introduced to make the analogy with molecular dynamics even more complete and corresponding to the velocities at which the $\Q$-tensor components change)} of the degrees of freedom and adaptively changes the size of the time step itself based on the behavior of the force and velocity during the most recent update. \DMS{For convex optimization problems the addition of inertia can be proven to enhance convergence \cite{polyak1964some}, although for more complex energy landscapes in general little can be proven.} Thus, while it is a heuristic approach, FIRE has been shown to be competitive with (or even faster than) conjugate gradient minimization \citep{bitzek2006structural,sheppard2008optimization,herbol2017computational}, all while maintaining an extremely light additional memory footprint and being highly  amenable to parallelization across multiple cores or multiple GPU units. \DMS{Note that FIRE was originally developed with atomistic simulations in mind, but it is increasingly being used more generally, including in the solution of PDEs \cite{zhou2019solution} and in machine learning applications \cite{wang2019search}. By the straightforward mapping mentioned above we are able to directly apply the pseudocode presented in Algorithm 1.}

\begin{algorithm}[H]
Initialize $\Q$-tensors at each lattice site, set \DAB{velocities} $v_i$ to zero\;
 \While{Minimization criteria not satisfied}{
  Update $q_i(\vec{x}_j)$, force $=-\nabla \mathcal{F}$, and $v_i$ using a velocity Verlet step\;
  Calculate power, $P$, as the dot product of the force and velocity vectors\;
  Rescale velocity by a parameter $\alpha$ which sets the inertia of the degrees of freedom\;

  \eIf{$P>0$}{
  	  \If{ $P$ has been positive for more steps than a threshold, $N_{min}$}{Increase the time step size and increase $\alpha$.}
}{
   Decrease the time step size, reset velocities to zero, reset $\alpha$ to initial value\;
  }
 }
 \caption{Pseudocode for FIRE minimization \citep{bitzek2006structural} }
\end{algorithm}

\begin{figure}[h!]
\begin{center}
\includegraphics[width=10cm]{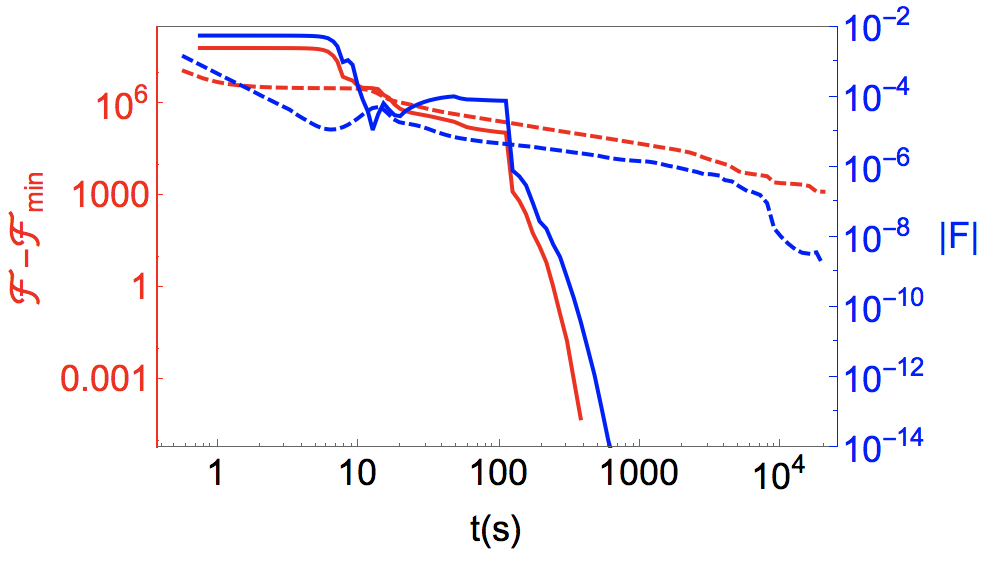}
\end{center}
\caption{ (Red) Energy relative to the uniform texture with preferred nematic order, $\mathcal{F}-\mathcal{F}_{min}$, and (blue) the norm of the residual force vector, $|F|/\sqrt{N}$, for bulk nematic (\DAB{l}attice size is $N=250^3$), starting from a randomly initialized configuration, as a function of wall-clock time. Solid lines are minimizations using FIRE  and dashed lines are those using gradient descent. As described in the text, we have tuned the minimization parameters (step size, etc.) for each algorithm separately and use identical hardware to make a one-to-one comparison.}\label{fig:minimizationTiming}
\end{figure}

We first demonstrate this efficient minimization in Fig. \ref{fig:minimizationTiming}, where we compare the system energy and average norm of the force on the degrees of freedom during the minimization of a lattice of $N=250^3$ sites in a cubic geometry with periodic boundary conditions. To make a fair comparison, we have performed both a FIRE and a gradient descent (GD) minimization on the same system using separately tuned minimization parameters for each algorithm. We use the same hardware for each simulation, and report the minimization progress in terms of the wall-clock time taken. Although it is sometimes common to report efficiency in such comparisons in units of function calls, for algorithms with very different numbers of arithmetic operations (each FIRE iteration requires more than twice the number of arithmetic operations compared to GD) such comparisons are often misleading.

\begin{figure}[h!]
\begin{center}
\begin{minipage}{9cm}
(a) \\
\includegraphics[width=\textwidth]{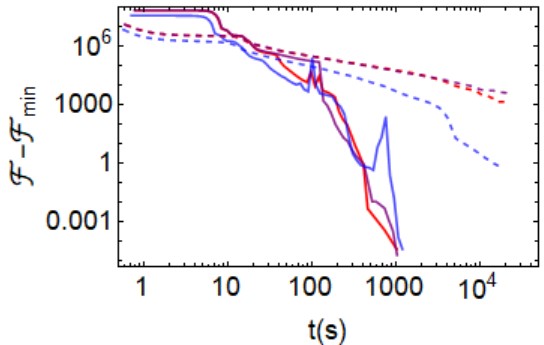}
\end{minipage}
\begin{minipage}{7.5cm}
(b) \\
\includegraphics[width=\textwidth]{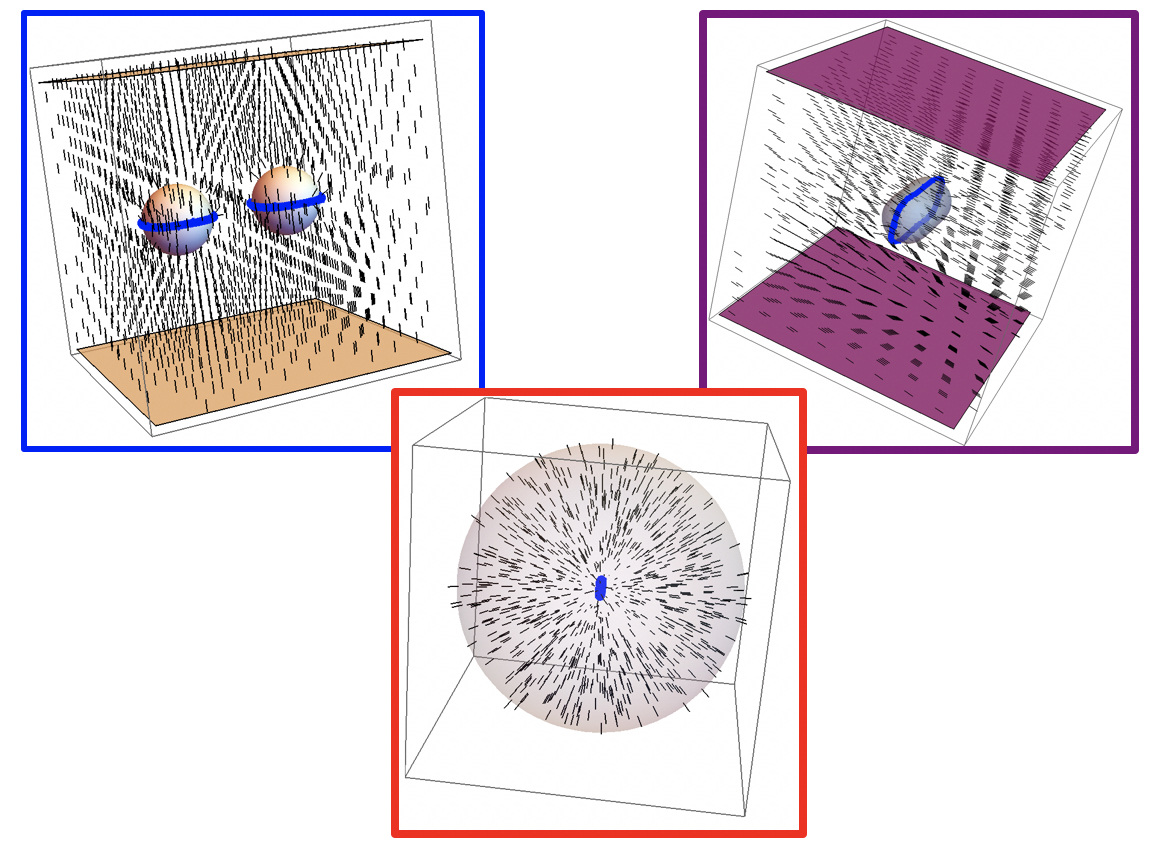}
\end{minipage}
\end{center}
\caption{(a) Energy relative to the minimized energy for three different geometries as a function of wall-clock time, in a lattice of size $N=250^3$ and starting from a randomly initialized configuration. As in Fig. \ref{fig:minimizationTiming}, solid lines are minimizations using FIRE  and dashed lines are those using gradient descent. As depicted in (b) showing the minimized configurations, the three sets of lines correspond to (Blue) two spherical colloids between parallel plates, all with homeotropic anchoring, (Red) the interior of a spherical droplet with homeotropic anchoring, and (Purple) a spherocylinder with homeotropic anchoring between parallel plates with planar degenerate anchoring. \DMS{These images were created using the ``multirankImages.nb'' Mathematica file included in the repository for making simple visualizations.}}\label{fig:objects}
\end{figure}

As Fig.  \ref{fig:minimizationTiming} makes clear, even in the trivial case of finding the uniform nematic ground state for a system with no boundary terms from a system initialized with random $\Q$-tensors at each lattice site, FIRE provides orders of magnitude improvement in the time taken to find minima. This performance of our default minimizer is not restricted to simple, bulk states of the liquid crystals. As we demonstrate in Fig. \ref{fig:objects} for a handful of simple (and well studied) arrangements of colloidal inclusions and boundaries, FIRE is very rapidly able to find these more complex minima, too. \DMS{As with any non-convex optimization solver, though, no guarantees are made by FIRE about avoiding particular local minima in favor of a true global minimum. Where this is a concern, we adopt the standard approach of minimizing from multiple different random initializations.} Particularly when coupled with a GPU, the substantial acceleration of FIRE-based minimizations enables the usefulness of the GUI, as the evolution of defect structures in response to user-instigated changes can be seen in real time.

Although numerical simulations of this size have been commonly used to make contact with experiments, in single-core operation it is impractical to simulate lattices much larger than $N \sim 300^3$, with the limiting factor being the wall-clock time required for CPUs and memory constraints for GPUs. Given a simulation with $\mathcal{N}$ degrees of freedom and spreading the work across $P$ processing units (either GPUs or CPUs), achieving ideal $\mathcal{N}/P$ scaling requires both low-latency communication between processors and algorithms that are themselves linear in $\mathcal{N}/P$. Fortunately, lattice-based models with only nearest- and next-nearest-neighbor interactions are trivial to parallelize using a pattern common to, e.g., spin glasses \citep{lulli2015highly}. \DMS{We use a standard spatial decomposition of the total number of lattice sites into rectilinear sub-regions (typically cubes, although other spatial partitions are easily implemented, and may be preferable for some simulation geometries). Each processing unit is assigned to one of these subregions, and is responsible for controlling and updating the lattice sites in that subregion. It also maintains information about the state of the ``halo'' of lattice sites that are neighbors, nearest-neighbors, and next-nearest neighbors of lattice sites at the boundary of the subregion it controls. Standard OpenMPI protocols \citep{gabriel2004open} are used during each simulation step to communicate information about the state of these halo sites to and from each processing unit in optimized sequences of uni-directional transfers.}

We now assess how our method's efficiency scales as the problem size is increased. Although \emph{strong scaling} (Amdahl's law) -- in which the total problem size is kept fixed and $P$ is increased -- is often important, it is well-established that the structure of the near-neighbor lattice interactions we simulate is embarrassingly parallel. Our real aim is to scale up the problem size itself and use many processors to simulate lattices that approach experimental scales. As such, \emph{weak scaling} (Gustafson's law) -- in which the amount of work per processing unit is kept constant -- is the relevant test. 

\DMS{One challenge to mention here is that when targeting energy minima -- as opposed to simply advancing a molecular dynamics simulation for a fixed number of time steps -- the number of minimization steps itself grows with the total system size. In general the convergence properties of different minimizers in non-convex settings are highly nontrivial. For simple geometries we are able to numerically probe this scaling  -- for instance, we find that in the absence of any boundary the number of minimization steps to achieve a target small force tolerance scales with the linear size of the system, whereas in the presence of a spherical colloid it scales roughly with $L^{3/2}$. In general, though, the approximate scaling may be hard to ascertain (and may depend on the target threshold for declaring a configuration to be in a minimum).}

\begin{figure}[h!]
\begin{center}
\includegraphics[width=9cm]{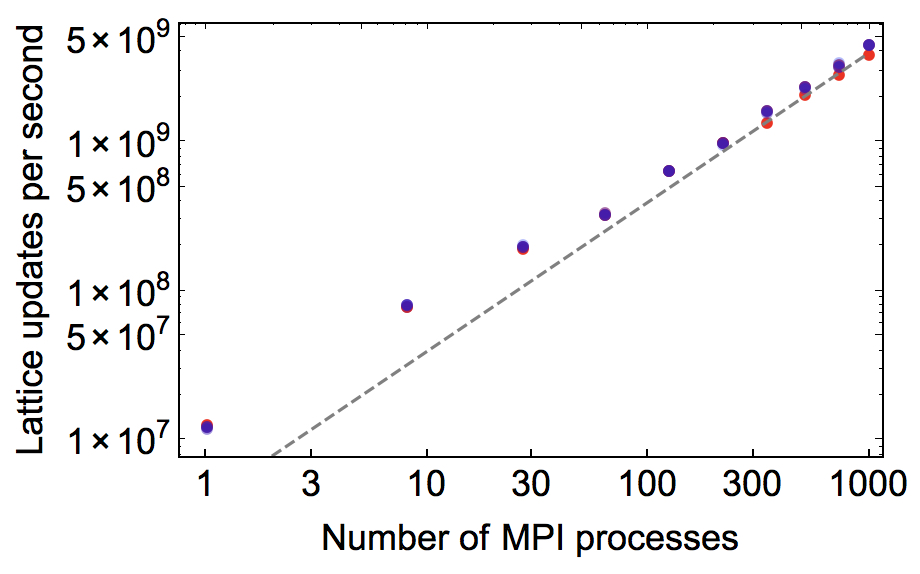}
\end{center}
\caption{Weak scaling performance of \PackageName\ on Comet, in total number of lattice site updates (i.e., (time steps)$\times$(ranks)$\times$($N_p$)) per second vs.\ the number of CPU processes, $P$, for a constant number of lattice sites per process. The points from dark red to light blue correspond to $N_p=75^3,\ 100^3,\ 125^3,\ 150^3,\ 250^3$ lattice sites per rank. The dashed gray line corresponds to ideal $\propto P$ scaling.}\label{fig:scaling}
\end{figure}

Turning instead, then, to the per-minimization-step timings, we present the weak scaling performance of \PackageName\  in Fig. \ref{fig:scaling}, where we compute the total number of lattice-site updates (i.e. $N$ times the number of simulated time steps) during a minimization in which we fix $N_p$, the number of lattice sites per processing unit, at several values and vary $P$. Consistent with a globally cubic simulation, we parallelized across $P=1^3,\ 2^3,\ 3^3,\ 4^3,\ 5^3,\ 6^3,\ 7^3,\ 8^3,\ 9^3,\ 10^3$ processors on the Comet XSEDE cluster, and studied computational performance for $N_p=75^3,\ 100^3,\ 125^3,\ 150^3,\ 250^3$. As expected, there are systematic drops due to increased communication costs as one goes from 1 core to multiple cores to multiple nodes, but \PackageName\  recovers ideal linear scaling of lattice updates with $P$ as $P$ grows very large. Additionally, there is a systematic degradation of performance for \emph{small} $N_p$, since in that case there is a more unfavorable ratio of halo sites to controlled sites for each processor.

Note that when we set the characteristic lattice spacing to correspond to $4.5$nm, the largest system simulated in this study, $N_p\times P = (250^3)\times10^3$, corresponds to a simulation domain of volume \DAB{1424 $\mu$m$^3$}. 

\section{Sample studies} \label{sec:SampleStudy}

\subsection{Companion defects to homeotropic spherical colloids}
In this section we apply \PackageName\ to the question of whether a hyperbolic hedgehog or a Saturn ring disclination loop provides the minimum-energy form of the topological companion defect to a homeotropic spherical colloid. As mentioned above, a larger colloid radius $a$ favors the dipolar configuration with a hedgehog, whereas smaller $a$ favors the quadrupolar configuration with a Saturn ring. As a result, the common rescaling of experimental dimensions to smaller $a/\xi_N$ in numerical modeling risks obtaining qualitatively different topological defect configurations. Besides increasing the simulation box size, altering the modeled material constants can restore qualitative agreement between experiment and simulation.  Here we explore the issue in detail, using \PackageName\ to systematically investigate the stability of hedgehogs relative to Saturn rings over a range of sizes and material parameters. 

The dipolar configuration with a hyperbolic hedgehog is the ground state for homeotropic colloidal particles near or above the micron scale \citep{poulin1997novel}. Terentjev's prediction of the alternative quadrupolar director field configuration with a Saturn ring disclination loop \citep{terentjev1995disclination} can be stabilized for large particles by confinement or external fields \citep{gu2000observation,loudet2001application}. Stark  \citep{Stark2001} demonstrated numerically using the Frank-Oseen free energy that the Saturn ring becomes metastable relative to the dipole for $a \lesssim$ 720 nm, with a defect core size $r_c=$ 10 nm. For $a \lesssim$  270 nm, the Saturn ring becomes the global ground state.

While the elastic energies of the two configurations are complicated to express, the Saturn ring is additionally penalized by a simple core energy per unit length, or line tension, $\gamma = \pi K/8$  \citep{Stark2001,deGennes_book}.  Because the Saturn ring maintains a radius $r_d$ just slightly larger than that of the colloidal particle, $r_d\approx 1.1 a$ \citep{Stark2001},  the total defect core energy penalty $E_c=2\pi r_d \gamma\propto K r_d$  of the Saturn ring grows linearly with the colloid radius. In contrast, the hyperbolic hedgehog has no  defect core dimension growing in size with the colloidal particle, helping to stabilize the dipole over the Saturn ring at larger colloid sizes. 

In order to numerically model multi-particle configurations in the dipolar size regime -- if we cannot exploit crystal symmetries to obtain a small unit cell \citep{Skarabot2008,nych2013assembly} -- we must either scale up the simulation volume to larger lattices, or stabilize the dipole at smaller particle sizes. We can achieve the latter by altering the materials constant ratios $\tilde B \equiv B/A$, $\tilde C \equiv C/A$ in Eq.~\ref{eq:fphase}. Together, these two ratios determine $S_0$ via Eq.~\ref{eq:S0def}, as well as the nondimensionalized free energy density of the nematic ground state $\tilde f_0 \equiv f_0/A$ with the energy well depth $f_0$ defined as in Eq.~\ref{eq:f0}.

By varying $\tilde B$ and $\tilde C$ such that $S_0$ remains fixed, we alter  the energetic cost per unit volume of melted nematic order in defect cores, $|f_0|$. The defect core size $r_c$ varies with the nematic correlation length $\xi_N$, which, from Eq.~\ref{eq:xiN}, scales as $\sim \sqrt{L_1/|f_0|}$. Thus, an increase in $|f_0|$ implies a decrease in the defect core size, which means effectively that the ratio $a/r_c$ of the particle size to the defect core size is increased without changing the size of the simulation lattice. The dipolar configuration is therefore expected to remain stable at smaller particle sizes. This technique was used in Ref.~\cite{luo2018tunable} to model a dynamical transition from Saturn ring to dipole as a colloidal particle approaches an undulated boundary, at simulation box sizes up to 50 times smaller than the experimental dimensions. 

\begin{figure}[h!]
\begin{center}
\includegraphics[width=7.5cm]{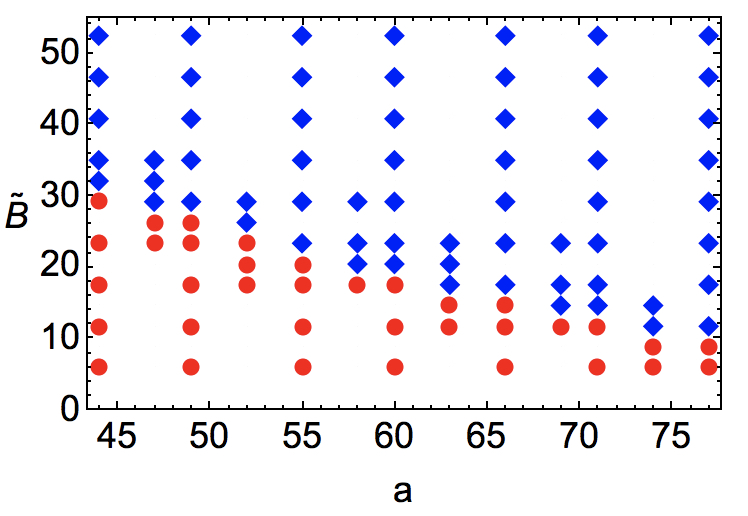}\includegraphics[width=7.5cm]{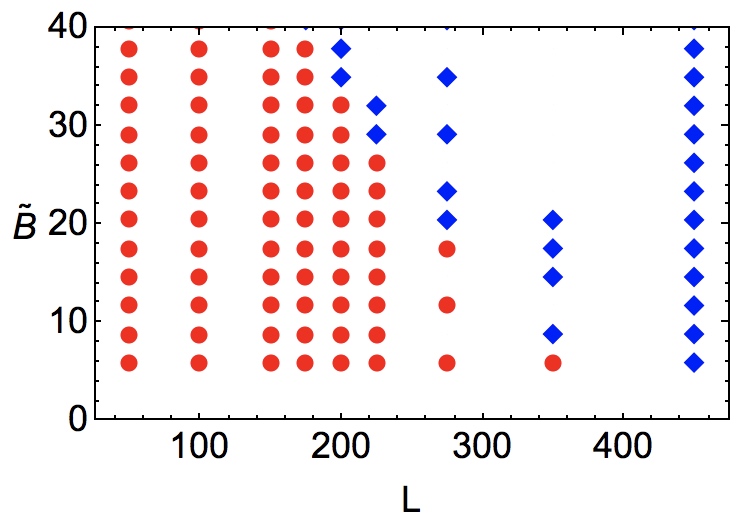}
\end{center}
\caption{Stability of dipolar defects around a spherical colloidal inclusion at fixed $S_0=0.53$ as a function of dimensionless \DAB{bulk free energy density} constant $\tilde{B}$, colloid size $a$, and linear system size $L$,  with $a$ and $L$ in units of the lattice spacing. Regions of parameter space with meta-stable dipolar configurations are shown with blue diamonds, stable quadrupolar configurations are shown with red circles. (Left) For fixed lattice size of $N=275^3$, we vary the dimensionless \DAB{bulk free energy density} constant $\tilde{B}$ and the colloidal radius $a$. (Right) For fixed ratio of colloidal radius to linear system size, $a=0.22\times L=0.22\times N^{1/3}$, we vary $\tilde{B}$ and $L$. Particularly for the larger values of $a$, one can see dependence of hedgehog meta-stability on $L$, indicating the importance of far-field distortions and boundary conditions (here, periodic). }  
\label{fig:hedgehogStability} 
\end{figure}

The results of this study are shown in Fig. \ref{fig:hedgehogStability}, which we parameterize by varying $\tilde{B}$ at fixed $S_0=0.53$ (i.e., setting $\tilde{C}=(2-\tilde{B} S_0)/(3S_0^2)$), along with the size of the spherical colloid and the lattice size. We test the stability of hyperbolic hedgehogs by initializing the surrounding lattice sites in the dipolar defect configuration suggested by Ref. \citep{Lubensky1998}, performing an energy minimization, and testing whether the resulting configuration has remained in the hedgehog state or transitioned to a Saturn ring configuration (thus, testing the meta-stability of the dipolar defect state as a function of system parameterization). At the values $\tilde B \approx 12$, $\tilde C \approx -10$ commonly used in modeling of 5CB \citep{ravnik2009landau}, we  find that the lower limit of hedgehog metastability is $a \approx 74$ lattice spacings, or about 330 nm. In this sample study we have imposed a large but finite anchoring strength at the colloid's surface. Weaker anchoring strength will affect the results, with a ``surface ring'' configuration replacing the dipole at low anchoring strength \citep{Stark2001}. 

\onecolumngrid

\twocolumngrid

We have also tested the meta-stability of the quadrupolar defect configuration by initializing the system in a Saturn ring configuration and minimizing, but we have not observed the spontaneous appearance of hedgehog defects from such simulations, indicating at least the meta-stability (if not absolute stability) of Saturn rings over the entire parameter range studied here. In addition to the effect of defect core size mentioned above, slight deviations in hedgehog meta-stability as a function of lattice size at fixed $\tilde{B}$ and $a$ seen in Fig. \ref{fig:hedgehogStability} indicate the importance of far-field distortion terms on the (meta-) stability of defect configurations.

\subsection{Patterned boundary conditions}

\DAB{To demonstrate the modeling of patterned boundaries in \PackageName, we examine a square array of alternating $\pm 1$ disclinations imprinted as a spatially varying anchoring direction on a planar substrate. Such an array was created experimentally by the authors of Ref.~\cite{murray2014creating}, by scribing lines into a polyimide surface with an atomic force microscope. As in that experiment, we give the opposing surface degenerate planar anchoring. In \PackageName, these boundary conditions are specified at each boundary lattice site through a user-prepared text file} \DMS{(see Sec.~\ref{sec: GUI} below).} \DAB{ We employ periodic boundary conditions  in the horizontal directions, and the anchoring strength $W$ at both surfaces is set to make the extrapolation length $K/W$ roughly equal to the lattice spacing. }

\begin{figure}[h!]
\begin{center}
\begin{minipage}{0.9\textwidth}
\raggedright
(a) \\
{\centering \includegraphics[width=\textwidth]{{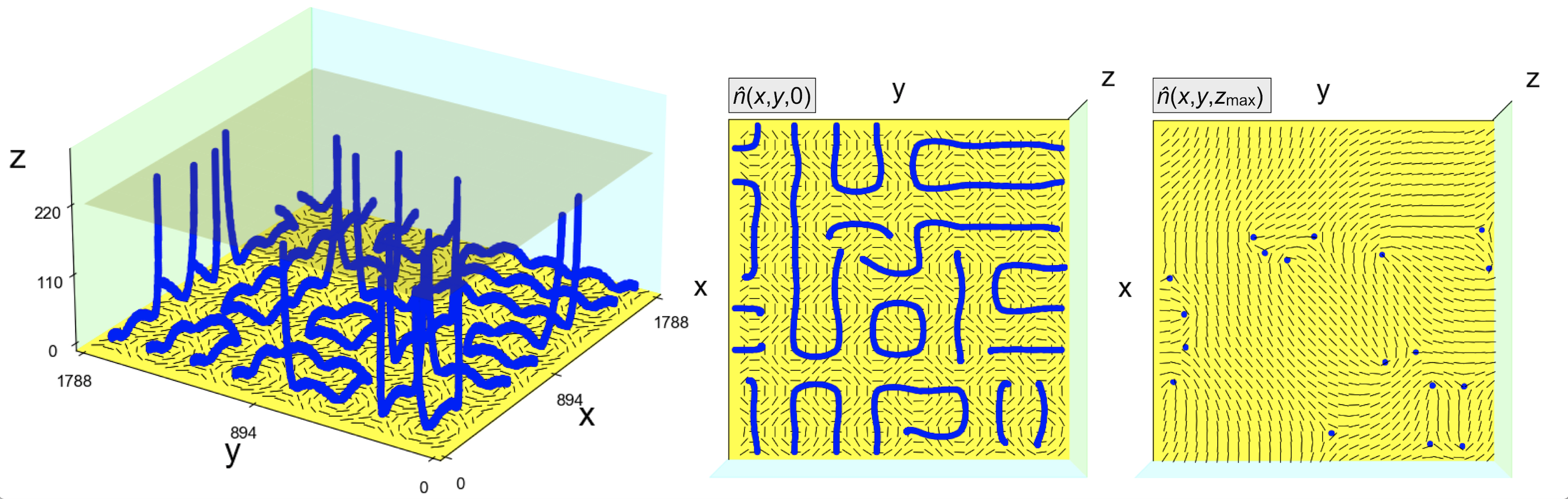}} }\\
(b) \\
{\centering \includegraphics[width=\textwidth]{{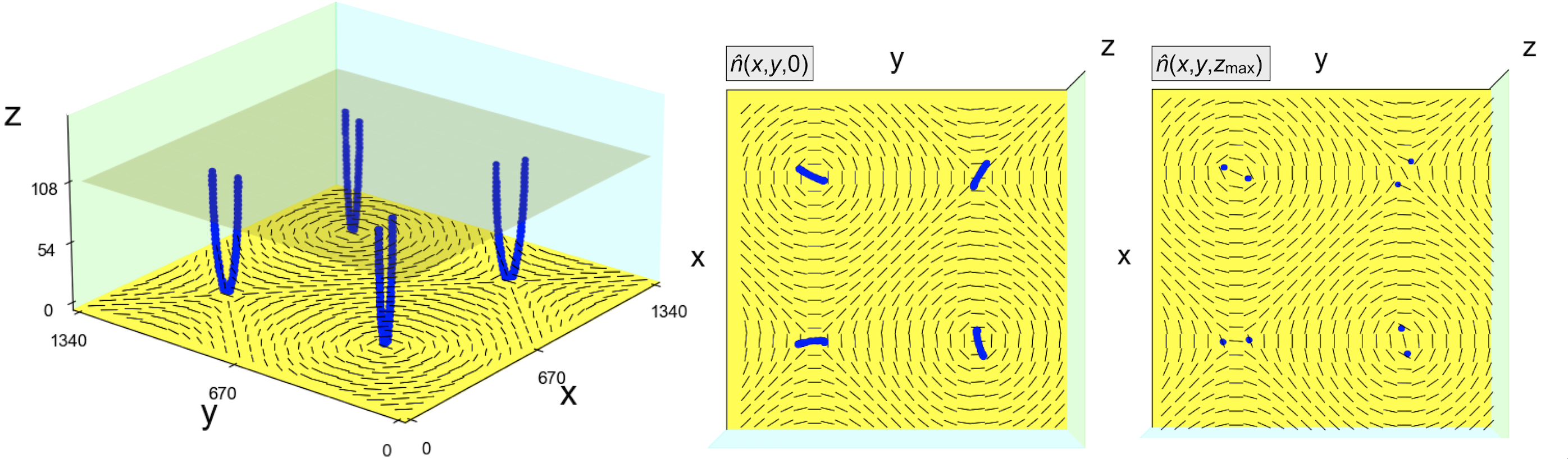}} } \\
(c) \\ 
{\centering \includegraphics[width=\textwidth]{{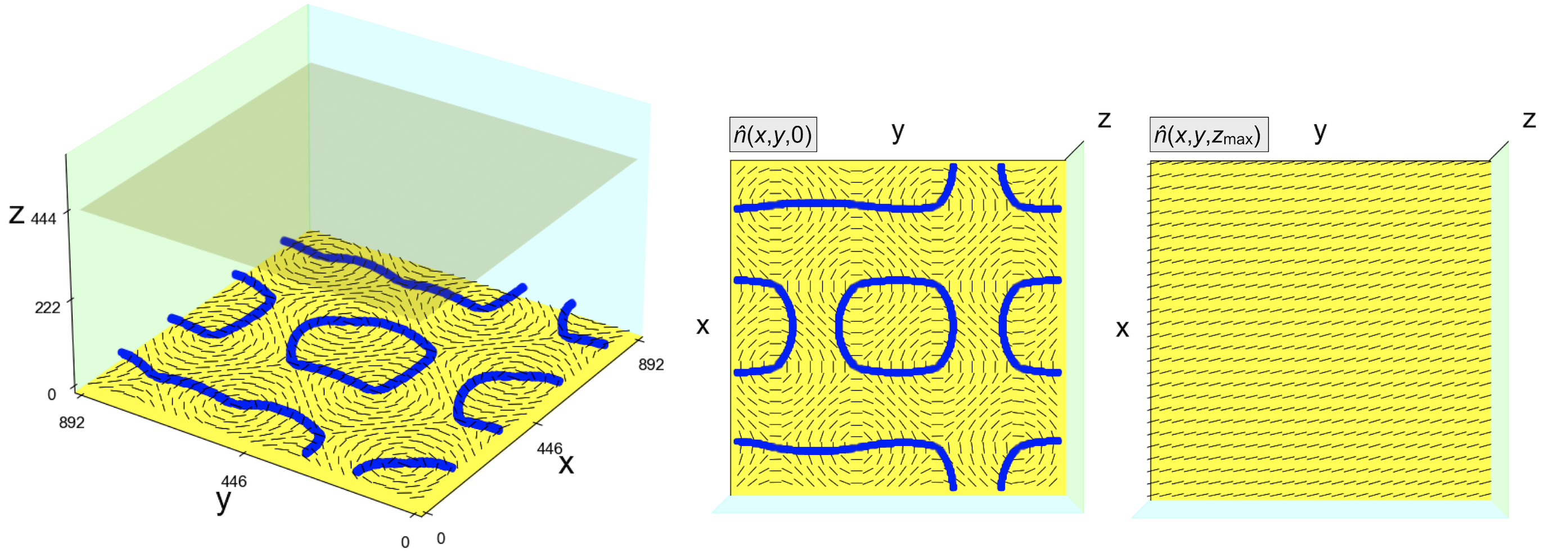}} }
\end{minipage}
\end{center}
\caption{\DAB{Numerically computed disclination configurations near planar substrates patterned with square arrays of alternating $\pm 1$ surface disclinations. The opposite planar boundary (transparent square) has degenerate planar anchoring. (a) An 8x8 array of surface disclinations with spacing equal to the cell thickness. (b) A 2x2 array of surface disclinations with spacing equal to six times the cell thickness. (c) A 4x4 array of surface disclinations with spacing equal to half the cell thickness. Configurations in (a)-(c) are partially energy-minimized. Disclinations are colored blue. Axes values are given in units of the lattice spacing. In each row, the second panel shows a top view of the disclinations in bulk and the director field in the plane of the patterned substrate; the third panel shows a top view of the director field on the opposite surface, along with the half-integer disclination points (if any) in that surface. These images are made using the ``visualize.py'' Python script included in the repository for taking saved configurations and making simple visualizations from the command line.  }}
\label{fig:patternedbdy}
\end{figure}

\DAB{Fig.~\ref{fig:patternedbdy}a shows the result} \DMS{of minimizing a cell of thickness $h =224$ lattice spacings,} \DAB{ corresponding to $\approx 1$ $\mu$m for 5CB,  and a spacing $d$ between defects equal to $h$.} \DMS{ We create an 8 by 8 array of defects,} \DAB{ so the total volume modeled is 64 $\mu$m$^3$, larger than the maximum size achievable with single core minimizations on a typical CPU ($\approx 10-20$ $\mu$m$^3$). Simulating several unit cells of the substrate patterning in this way allows us to observe a labyrinthine configuration of half-integer disclination lines near the plane of the substrate, connecting neighboring surface-defects. Meanwhile, some disclination lines are vertical, traveling between the two surfaces and imprinting a $+\frac{1}{2}$ or $-\frac{1}{2}$  defect profile on the top surface. } \DMS{The stopping condition for the minimization here was a somewhat modest force tolerance, allowing these large-system-size studies to be completed in less than 24 hours. } \DAB{While clearly not completely equilibrated, the horizontal disclination labyrinth is similar to a domain wall texture observed experimentally  in Ref.~\cite{murray2014creating}, which may also be kinetically trapped. Absent from the texture in Fig.~\ref{fig:patternedbdy}a is the $\pm 1$ non-singular escaped configuration, which did appear in the experiments.}

\onecolumngrid

\twocolumngrid

\DAB{The energetic cost per unit length of disclination lines implies that the vertical configuration is favored by smaller cell thickness $h$. Indeed, }\DMS{as shown in Fig.~\ref{fig:patternedbdy}b, } \DAB{when we decrease $h/d$ from 1 to $\frac{1}{6}$, only vertical disclinations appear, in pairs of $+\frac{1}{2}$ or $-\frac{1}{2}$ disclinations from the ``splitting'' of the $\pm 1$ surface defects. This defect splitting was  sometimes observed in Ref.~\cite{murray2014creating} in place of the escaped configuration. Conversely, as shown in Fig.~\ref{fig:patternedbdy}c, only horizontal disclinations appear when $h/d$ is increased to 2. Extensions to even larger defect arrays, to curved boundaries, and to spatially nonuniform anchoring types can be explored in the same manner in \PackageName.   }

\section{Rapid prototyping with GUI interface} \label{sec: GUI}
Figure \ref{fig:gui} shows a screenshot of the graphical user interface in action, and the \href{https://www.youtube.com/watch?v=4VPDqSytuZc}{supplemental video} \DMS{and accompanying narrative transcript of the video in the supplemental text} shows a representative demonstration of its use. Here we discuss some of its current functionality. Initialization dialog boxes 
 allow the user to set the simulation size, the computational resource to use (CPU or GPU, autodetecting whether CUDA-capable resources are available for use), and parameters for the bulk and distortion free energy density. This generates a random bulk configuration of $\Q$\DAB{-}tensor lattice sites with periodic boundary conditions. For the visualization pane the user can specify the density and magnitude of directors to draw (taken to be the direction of the largest eigenvector of $\Q$ at each site), and can freely rotate and zoom in on the configuration, as well as highlight in blue defects defined locally by regions where the largest eigenvalue of $\Q$ falls below some threshold.

In the top left are buttons allowing the user to specify parameters from one of two energy minimization techniques to use (FIRE and Nesterov's Accelerated Gradient Descent); the resulting dialog boxes are populated with values that we typically find to be efficient for default parameter choices in the bulk and distortion energies, although some amount of tuning may be quite beneficial (particularly when changing the distortion terms $L_2$ through $L_6$). The ``Minimize'' button performs the requested energy minimization (either until a target force tolerance is attained or the maximum number of iterations is reached), with the option to visualize the results only at the end or to watch the minimization proceed. The ``File'' dialog box allows the currently visualized state of the system to be saved for separate analysis or processing.

Note that menu items allow any of the terms in the energy functional governing the simulation, Eq. \ref{eq:Fldg}, to be changed at any moment. This allows, for example, the user to first minimize a system with some values of the distortion constants and then perform repeated minimizations as those values are changed, observing the stability or metastability of defect structures as this is done.

Two buttons allow the user to introduce boundaries and colloidal inclusions into the system. ``Simple'' objects are spheres and flat walls with either normal homeotropic or degenerate planar anchoring conditions. Arbitrarily complex boundary conditions (taking any shape, with degenerate planar and homeotropic anchoring conditions not restricted by the direction of the surface normal) can be added by preparing a simple text file that the program can read in -- an example script that generates the custom boundary file used in Fig. \ref{fig:gui} is included in the ``/tools'' directory of Ref. \cite{landauDeGUI}.

With boundaries and colloids (``objects'') in place, \DMS{some manipulations of these objects are accessible via drop-down menus.} The positions of these objects within the simulation can be directly modified, so the user could place an ellipsoidal particle, perform a minimization, change the position, re-minimize the system, and record the different energy minima attained. We include an option to automate this type of operation (which can be used to build up the potential of mean force from the liquid crystal and colloid interactions) for convenience. A near-term addition will be  allowing objects to move according to the integrated stresses at their surface (or according to the energetic results of various trial moves); the user will then be able to separately ``Minimize'' just the liquid crystal sites or ``Evolve [the] system'' by allowing both liquid crystalline and colloidal degrees of freedom to change simultaneously.

Finally, to facilitate moving from GUI prototyping to larger-scale MPI studies, we have included the ability to record system initialization and sequences of commands entered in the graphical user interface, and then save this sequence of commands as a new file that can be separately compiled and executed in non-GUI operation. This file has its own set of command line options, primarily so that it can be made to work as an MPI executable and so that the system size of the simulation it represents can be rescaled to a larger value. We highlight this GUI-prototyping approach as a visual alternative to the scripting-language approaches of molecular-dynamics packages like LAMMPS \citep{plimpton1995fast} or HOOMD-blue \citep{anderson2008general} for specifying complex sequences of system initialization, energy minimizations, and the introduction of objects, fields, and boundary conditions. We believe that this seamless visual-prototyping-to-MPI-scalable pipeline will be beneficial to researchers interested in accessing experimental-scale simulations.

\section{Discussion} \label{sec: Discussion}

As demonstrated in our sample study, \PackageName\ utilizes MPI to enable LdG modeling at typical size scales of  experimental relevance, at the $\sim 10$ $\mu$m range, with fast convergence enabled by the FIRE algorithm. Besides the colloidal defect configurations \DAB{and patterned boundaries} discussed here, another immediate use is for the study of cholesterics, where typically fewer than ten pitches can fit inside a simulation box using a single processor, but using \PackageName\ tens of pitches can be modeled. While it may not be realistic at present to \textit{frequently} conduct simulations with $10^3$ processors, using \PackageName\ on computer clusters will facilitate demonstration of how numerical results scale with system size, allowing reasonable extrapolations to experimental scales.   

For modeling at the $\sim 1$ $\mu$m range or smaller, \PackageName's combination of FIRE with GPU computing offers a substantial speedup, enabling users to manipulate the simulated conditions in a GUI environment and observe the change in energy-minimized configurations. The GUI is useful for running ``real-time'' tests   of proposed configurations which can then be modeled at larger scales with MPI.

Likewise, the GUI will also be useful to experimentalists in quickly identifying more optimal properties of nematics, colloidal particles, boundaries, etc.\ in order to achieve targeted topological or self-assembled configurations. In general, numerical modeling can aid experimental studies not only in developing theoretical understanding of nematic structures and energy landscapes, but also in performing high-throughput searches through these design spaces. For example, geometric compatibility conditions favoring lock-and-key assembly of particles and patterned walls \citep{eskandari2014particlesoftmatter,luo2018tunable}, or particle design promoting assembly into photonic crystals, can be optimized more efficiently in numerics, to help guide the increasingly sophisticated uses of fabrication techniques  such as photolithography and two-photon polymerization \citep{martinez2014mutually}. 
An ambitious but important direction for future development is therefore to efficiently explore design parameter spaces in numerical modeling, possibly employing genetic algorithms and techniques from machine learning.

There are some near-term directions for future development of  \PackageName\ that we anticipate will increase the usefulness of this open-source software to the liquid crystals research community. 
 An expanded  library of $\Q$-initialization options will facilitate investigations of chiral liquid crystals, topologically entangled or knotted defect configurations \citep{Ravnik2007,Tkalec:2011lj,tasinkevych2014splitting,machon2014knotted}, and periodic defect arrays \DAB{\citep{murray2014creating,suh2019topological}}, for example. A major advance would be adding a flow field coupled to $\Q$ by Beris-Edwards nematodynamics, for investigations of microfluidic geometries and active nematics.

Incorporating motion of colloidal particles into the modeling is another area for useful developments. In the experimental system, energy is minimized not only over $\Q$ but also over the positions and (if applicable) orientations of colloidal particles. At present, \PackageName\ takes these latter degrees of freedom as input parameters, and a free energy landscape can be mapped either informally using the GUI or more systematically on a computer cluster. Thus one desired future improvement is to allow overdamped translation and rotation of colloidal particles within the program,  downhill in the energy landscape, based on trial moves or on estimated nematic elastic stresses felt by the particle  \citep{vskarabot2008hierarchical}. The trial move approach, requiring several re-minimizations of $\Q$ at each time step, is made less cumbersome by improved convergence speed of the FIRE algorithm. 

Finally, we hope that \PackageName's GUI interface will be useful in physics education. Interacting with  a fast and ``hands-on'' version of the numerical modeling, students at the undergraduate or beginning graduate level can quickly gain experience and intuition for liquid crystals. This will help to capitalize on the position of  liquid crystals as one of the most accessible, and visualizable, physical realizations of  abstract topological ideas relevant to many areas of physics.

\section*{Acknowledgments}
DAB thanks Gareth Alexander for introducing him to LdG numerical modeling, and gratefully acknowledges illuminating discussions with Simon \v{C}opar, Miha Ravnik, and Slobodan \v{Z}umer. DMS was supported by NSF-POLS- 1607416 as well as Simons Foundation Grant Number 454947. The Tesla K40s used for this research were donated by the NVIDIA Corporation. We acknowledge computing support via an XSEDE allocation on Comet through Grant No. NSF-TG-PHY190027, \DAB{and from the Multi-Environment Computer for Exploration and Discovery (MERCED) cluster at UC Merced, which was funded by National Science Foundation Grant No. ACI-1429783.}

\section*{Data Availability Statement}
The open source code described in this work can be found at Ref. \citep{landauDeGUI} and used to reproduce all data in the manuscript. Documentation for the software is maintained at \citep{landauDeGUIdocumentation}\DMS{, and can also be generated with doxygen from the source code.}

\bibliography{landauDeGUI_arxiv}

\newpage
\onecolumngrid

\appendix
\section{Description of demonstration video}
In this appendix we provide a narrative transcript of how the graphical user interface is being manipulated in the accompanying video for this manuscript (\href{https://www.youtube.com/watch?v=4VPDqSytuZc}{youtube link}).

\subsection{System initialization (0:00 - 0:12)}
After executing the openQminGUI.out command a brief splash screen is followed by a ``System initialization'' screen. From this screen the number of lattice sites in the $\hat{x}$, $\hat{y}$, and $\hat{z}$ direction to simulate can be separately specified, or for convenience (when simulating cubic domains) all three can be controlled at once by changing the dialog box under ``L.'' The constants determining the bulk free energy of the nematic can be individually set below these system size options. The program attempts to auto-detect any CUDA-capable GPU devices available, and the user can select which detected GPU to use (or whether to use only CPU resources) from the ``available compute units'' drop-down menu. 

The user can additionally specify which approximation to use for the distortion contribution to the free energy by cycling through the ``Number of constants'' dialog box. After clicking the ``Initialize'' button, a separate screen pops up allowing the user to specify the constants associated with the distortion energy chosen; these choices are finalized by clicking on the ``set constants'' button.

\subsection{Manipulating the graphical representation (0:12 - 0:25)}
After the system is initialized a graphical view of the current configuration appears. A single director, whose direction corresponds to the eigenvector of the Q-tensor with largest eigenvalue, is drawn for every $n$th lattice site in each direction, where $n$ is the number appearing in the ``lattice skip'' dialog box. The graphical view can be freely rotated by clicking and dragging on the image, with a slider allowing the user to zoom in and out, and the size of the director drawn at each site can be modified by changing the ``director scale.''

\subsection{Minimization of bulk systems (0:25 - 0:49)}
At present two different minimizers are included in the GUI -- one corresponding to the FIRE algorithm and the other to Nesterov's accelerated gradient methods. By default FIRE is chosen, but either can be selected by pressing the corresponding button on the left side of the screen (which brings up the set of parameters associated with the chosen minimization algorithm, all of which can be changed from their defaults by the user). Pressing the ``Minimize'' button executes the chosen minimization routine, which runs for a maximum number of iterations but which includes an early-stopping condition if the norm of the total force vector on all of the lattice sites falls below a user-specified tolerance. Pressing either the ``Minimize'' or ``Evolve system'' buttons will execute the same minimization command repeatedly, allowing the user to control the degree of minimization quite easily.

At any moment the user can, additionally, reset the system to random initial conditions (with either a reproducible or a non-reproducible scheme for generating the random numbers associated with the random initial conditions), allowing the user to explore the influence of initial conditions on the minimum found. If the ``display progress visually'' checkbox is activated, during the course of minimization the current state of the system will be periodically displayed in the visualization pane, allowing the user to watch the system as it descends in the energy landscape.

\subsection{Adding objects and progressive minimizations (0:49 - 2:31)}
The settings specifying the system initialization can be returned to at any time using the ``System setup'' $\rightarrow$ ``Reset the system'' menu command. 

\subsubsection{Adding simple boundaries (1:24 - 1:32)}
The ``Add a simple object'' button allows the user to easily add a limited number of predefined types of boundary conditions to the simulation. The user  can add a spherical colloid (specifying its position and radius) or a flat wall which has a normal in the $\hat{x}$, $\hat{y}$, or $\hat{z}$ direction and sits on a particular lattice plane. From the drop-down menu the user can specify either  homeotropic and degenerate planar anchoring conditions of a desired strength.

\subsubsection{Changing forces on the fly (1:32 - 2:31)} 
Both the bulk and distortion free energy constants can by changed by clicking through the ``Change forces'' menu buttons. This allows the user to specify complex minimization pathways with ease, and it may be useful in preparing unusual types of initial conditions. For instance, it may be beneficial to minimize to a target state by slowly varying the distortion terms by only gradually adding in degrees of chirality, etc. In the present example, the chiral state with preferred wavenumber $q_0$ is more easily found by first minimizing in the simple one-constant approximation and then moving to a more complicated set of distortion terms.

For visualization purposes, the checkboxes to the right of the visualization pane allow the user to see individual lattice planes of director configurations in each of the primary lattice directions. This reduced representation of the system often makes it much easier to parse the overall structure of the directors throughout the simulation domain.

\subsection{Watching defects evolve (2:31 - 3:24)}
After resetting the system to a cubical domain and redrawing the visualization pane, planar boundaries are added and a minimization quickly finds the (simple) ground state of the system.  Two homeotropic spheres are added, guaranteeing that in the ground state there must be defects in the director configuration. By clicking on the ``show defects'' checkbox, lattice sites whose largest eigenvalue falls below the specified ``defect threshold'' are represented not with director lines but with blue spherical dots in the visualization pane. The evolution of the defect structure can be readily visualized in real time while the minimization proceeds.

\subsection{Adding completely customized boundary conditions (3:24 - 4:19)}
In the likely case that the user wants to go beyond arrangements of planes and spheres as boundary conditions in the simulation, a separate button allows the user to ``Add [an] object from [a] carefully prepared text file.'' This file must, indeed, follow a very specific format (described in the ``Preparing a custom boundary file'' section of the README.md included in the code repository), but it allows the user to specify any number of separate objects, each of which may be composed of a completely arbitrary set of lattice sites and where each lattice site can have an arbitrary planar or homeotropic boundary free energy (not limited, for instance, to using a surface normal to define directions of preferred anchoring). 

Whereas predefined visualization options can be used for the simple spheres and planes that are easy to add, turning off the ``built-in boundary visualization'' checkbox represents every object site as a white semi-transparent sphere. The particular example shown in the video corresponds to a substrate topography whose structure was taken from a United States Geological Survey of Mt. Katahdin (a prominent mountain in the northeastern US), with a pyramidal colloidal inclusion placed above the substrate. The ``almost-smiley-face'' pattern of the defect structure induced by this set of boundary conditions forms the unofficial logo of the \PackageName\  software package.

\end{document}